\documentstyle[epsfig]{czjphys2}

\begin{document}
\title{HOW COULD THE PROTON TRANSVERSITY BE MEASURED}
\authori{A.V. Efremov\,\footnote{E-mail: efremov@thsun1.jinr.ru}}
\addressi{Joint Institute for Nuclear Research, Dubna, 141980 Russia}
\authorii{}     
\authoriii{}     
\headtitle{How could the proton transversity be measured }
\headauthor{A.V. Efremov}  
\specialhead{A.V. Efremov: How could the proton transversity be
measured.}
\evidence{A}  \daterec{XXX}    
\cislo{0}  \year{2000}
\setcounter{page}{1}
\pagesfromto{000--000}

\maketitle

\begin{abstract}
The perspectives of two new nonstandard methods of transversal quark
polarization measurement are considered: the jet handedness and the
so-called "Collins effect" due to spin dependent T-odd fragmentation
function responsible for the left-right asymmetry in fragmenting of
transversally polarized quarks. Recent experimental indications in favor
of these effects are observed:

1.The correlation of the T-odd one-particle fragmentation functions
found by DELPHI in $Z\to 2$-jet decay. Integrated over the fraction of
longitudinal and transversal momenta, this correlation is of 1.6\% order,
which means order of 13\% for the analyzing power.

2.A rather large ($\approx10\%$) handedness transversal to the production
plane observed in the diffractive production of ($\pi^-\pi^+\pi^-$)
triples from nuclei by the $40\,{\rm GeV/c}$ $\pi^-$--beam. It shows a
clear dynamic origin and resembles the single spin asymmetry behavior.

All this makes us hope to use these effects in polarized DIS  experiments for
transversity measurement. The first  estimation of transversity was done by 
using the azimuthal asymmetry in semi-inclusive DIS recently measured by
HERMES and SMC.
\end{abstract}

\section{Introduction}

My talk concerns recent progress in the possibility of transverse quark
polarization measurement. This is interesting in many aspects, and one of the
most important of them is quarks transversity distribution in proton $h_1(x)$
measurement. Let me recall that there are three most important (twist-2) parton
distribution functions (PDF) in a nucleon: a non-polarized distribution
function $f_1(x)$, longitudinal spin distribution  $g_1(x)$ and transversal
spin distribution  $h_1(x)$. The first two have been more or less successfully
measured experimentally in classical deep inelastic scattering (DIS)
experiments, but the measurement of the last one is especially difficult since
it belongs to the class of the so-called helicity odd structure functions and
can not be seen there. To do this, one needs to know the transversal polarization
of a quark scattered from a transversally polarized target.

There are several ways to do this:
\begin{enumerate}
\item
To measure the polarization of a self-analyzing hadron into which the quark
fragmentizes in a semi-inclusive DIS, e.g. $\Lambda$-hyperon\cite{augustin}.
The drawback of this method, however, is a rather low rate of quark fragmentation
into $\Lambda$-particle ($\approx 2\%$) and especially that it is mostly
sensitive to $s$-quark polarization.
\item
To measure a spin-dependent T-odd parton fragmentation function (PFF)
\cite{muldt,muldz,mulddis} responsible for the left-right asymmetry in
fragmentation of a transversally polarized quark with respect to quark
momentum--spin plane. (The so-called "Collins asymmetry" \cite{collins}.)
\item
To measure the transversal handedness in multiparticle parton
fragmentation \cite{hand}, i.e. the correlation of quark spin 4-vector $s_\mu$ and
particle momenta $k_\nu$, $\epsilon_{\mu\nu\sigma\rho}s^\mu k_1^\nu k_2^\sigma
k^\rho$ ($k=k_1+k_2+k_3+\cdots$).
\end{enumerate}

The last two methods are comparatively new, and only the last years some
experimental indications the the T-odd PFF and the transversal handedness have
appeared \cite{todd,prev,yaf99}. Last PRAHA-SPIN Conference I have
presented details of these experiments \cite{czjp99}. Now I shortly repeat
the result of T-odd PFF measurements (Sect. 2) and concentrates on
its application for estimation of the proton transversity distribution
(Sect. 3), using the recently measured azimuthal asymmetry in semi-inclusive
DIS. Sect. 4 is reserved for conclusions.

\section{T-odd quark fragmentation function}

The transfer of nucleon polarization to quarks is investigated in
deep-inelastic polarized lepton -- polarized nucleon scattering experiments
\cite{rep}. The corresponding nucleon spin structure functions for the
longitudinal spin distribution $g_1$ and  transversal spin distribution
$h_1$  for a proton are well known.  The {\it inverse} process, the spin
transfer from partons to a final hadron, is also of fundamental interest.
Analogies of $f_1,\ g_1$ and $h_1$ are functions $D_1,\ G_1$ and $H_1$,
which describe the fragmentation of a non-polarized quark into a
non-polarized hadron  and a longitudinally or transversely polarized quark
into a longitudinally or transversely polarized hadron, respectively
\footnote{We use the notation of the work \cite{muldt,muldz,mulddis}.}.

These fragmentation functions are integrated over the transverse momentum
$\mathbf{k}_T$ of a quark with respect to a hadron. With $\mathbf{k}_T$
taken into account, new possibilities arise. Using the Lorentz- and
P-invariance one can write, in the leading twist approximation, 8 independent
spin structures \cite{muldt,muldz}. Most spectacularly it is seen in the
helicity basis where one can build 8 twist-2 combinations, linear in spin
matrices of the quark and hadron {\boldmath$\sigma$}, $\mathbf{S}$ with
momenta $\mathbf{k}$, $\mathbf{P}$. Especially interesting is a new T-odd
and helicity even structure that describes a left--right asymmetry in the
fragmentation of a transversely polarized quark:
$
H_1^\perp\mbox{\boldmath$\sigma$}({\mathbf P}\times
{\mathbf k_T)}/P\langle k_T\rangle,
$
where the coefficient $H_1^\perp$ is a function of the longitudinal
momentum fraction $z$, quark transversal momentum  $k_T^2$ and
$\langle k_T\rangle$ is an average transverse momentum.

In the case of fragmentation to a non-polarized or a zero spin hadron, not
only $D_1$ but also the $H_1^\perp$ term will survive. Together with its
analogies in distribution functions $f_1$ and $h_1^\perp$, this opens a
unique chance of doing spin physics with non-polarized or zero spin
hadrons! In particular, since the $H_1^\perp$ term is helicity-odd, it
makes possible to measure the proton transversity distribution $h_1$ in
semi-inclusive DIS from a transversely polarized target by measuring the
left-right asymmetry of forward produced pions (see~\cite{mulddis,kotz} and
references therein).

The problem is that, first, this function is completely unknown both
theoretically and experimentally and should be measured independently.
Second, the function $H_1^\perp$ is the so-called T-odd fragmentation
function: under the naive time reversal $\mathbf{P},\ \mathbf{k}_T,\
\mathbf{S}$ and {\boldmath$\sigma$} change sign, which demands a purely
imaginary (or zero) $H_1^\perp$ in the contradiction with hermiticity.
This, however, does not mean the break of T-invariance but rather the
presence of an interference term of different channels in forming the final
state with different phase shifts, like in the case of single spin
asymmetry phenomena~\cite{gasior}. A simple model for this function could
be found in~\cite{collins}. It was also conjectured~\cite{jjt} that the
final state phase shift can be averaged to zero for a single hadron
fragmentation upon summing over unobserved states $X$. Thus, the situation
here is far from being clear.

Meanwhile, the data collected by DELPHI (and other LEP experiments) give a
unique possibility to measure the function $H_1^\perp$.  The point is that
despite the fact that the transverse polarization of a quark ( an
antiquark) in Z$^0$ decay is very small ($\cal{O}(m_q/M_Z)$), there is a
nontrivial correlation between transverse polarizations of a quark and an
antiquark in the Standard Model:  $C^{q\bar q}_{TT}=
{(v_q^2-a_q^2)/(v_q^2+a_q^2)}$, which reaches rather high values at $Z^0$
peak: $C_{TT}^{u,c}\approx -0.74$ and $C_{TT}^{d,s,b}\approx -0.35$.  With
the production cross section ratio $\sigma_u/\sigma_d=0.78$ this gives the
value $\langle C_{TT}\rangle\approx -0.5$ for the average over flavors.

The spin correlation results in a peculiar azimuthal angle dependence of
produced hadrons (the so-called "one-particle Collins asymmetry"), if the
T-odd fragmentation function $H_1^\perp$ does exist~\cite{collins,colpsu}.
The first probe of it was done three years ago~\cite{delnote95} by using a
limited DELPHI statistics with the result
$\langle{H_1^{\perp}/D_1}\rangle\le 0.3$, as averaged over quark
flavors.

A simpler method has been proposed recently by an Amsterdam group
\cite{muldz}. They predict a specific azimuthal asymmetry of a hadron in a
jet around the axis in direction of the second hadron in the opposite jet
\footnote{ We assume the factorized Gaussian form of $k_T$ dependence
for $H_1^{q\perp}$ and $D_1^q$ integrated over $|k_T|$.}:
\begin{eqnarray}
{{\rm d}\sigma\over {\rm d}\cos\theta_2 {\rm d}\phi_1}\propto
(1+\cos^2\theta_2)\cdot \left(1+ {6\over\pi}\left[{H_1^{q\perp}\over
D_1^q}\right]^2 C_{TT}^{q\bar q}{\sin^2\theta_2\over
  1+\cos^2\theta_2}\cos(2\phi_1)\right)
\label{mulders}
\end{eqnarray}
where $\theta_2$ is the polar angle of the electron and the second hadron
momenta $\mathbf{P}_2$, and $\phi_1$ is the azimuthal angle counted off the
$(\mathbf{P}_2,\, \mathbf{e}^-)$-plane.  This looks simpler since there is no
need to determine the $q\bar q$ direction.

This analysis \cite{todd} covered  the DELPHI data collected from 1991
through 1995. Only the leading particles in each jet of two-jet events was
selected both positive and negative. The corrected for acceptance
histograms in $\phi_1$  for each bin of $\theta_2$ were fitted by the
expression~\footnote{ The term with $\cos\phi_1$ is due to the twist-3
contribution of usual one-particle fragmentation, proportional to the
$k_T/E$.} $P_0(1+P_2\cos2\phi_1 + P_3\cos\phi_1)$.
The $\theta_2$-dependence of $P_2$ in the whole interval of $\theta_2$
was fitted according to Expr. (\ref{mulders}) with the result
$$
P_2(\theta_2)=-(15.8\pm3.4){\sin^2\theta_2\over 1+\cos^2\theta_2}
{\rm ppm}
$$

The corresponding analyzing power, summed over $z$ and averaged over quark
flavors with $\langle C_{TT}\rangle\approx -0.5$ (assuming
$H_1^{\perp}=\sum_H H_1^{\perp, q/H}$ is flavor-independent)
according to Exp. (\ref{mulders}) is
\begin{equation}
\left|{\langle H_1^{\perp}\rangle\over\langle D_1\rangle}\right|
=12.9\pm1.4\% \ .
\label{apower}
\end{equation}
The systematic errors, however, are by all means larger than the statistical 
ones and need further investigation.

\section{Proton transversity estimation}

Recently azimuthal asymmetries in semi-inclusive hadron production on
longitudinally (HERMES \cite{avakdis99}) and transversally (SMC
\cite{bravardis99}) polarized targets where reported which together with
DELPHI result (\ref{apower}) allows to an estimation for transversity
distribution.

The T-odd azimuthal asymmetry in semi inclusive DIS $ep\to e'\pi^+X$ which
HERMES try to measure consist of two sorts of terms (see \cite{mulddis} Eq.
(115)): a twist-2 asymmetry $\sin2\phi_h$ and a twist-3 asymmetry
$\sin\phi_h$. The angle $\phi_h$ here is the azimuthal angle around
$z$-axis in the direction of virtual $\gamma$ momentum in the parton Breit
frame counted from the electron scattering plane. The first asymmetry is
proportional to the $k_T$-dependent transversal quark spin distribution in
a longitudinally polarized proton, $h_{1L}^\perp$, while the second 
contains two parts:  one term is again proportional to $h_{1L}^\perp$; and
the second, to the twist-3 distribution function $h_L$.

The experimentally observed $\phi$ dependence in HERMES data shows no
noticeable trace of $\sin 2\phi$ term. Thus, as a crude approximation one
can assume a smallness of $h_{1L}^\perp\gg h_L$. For the same reason $h_L
= h_1$ (see \cite{mulddis} Eq. (C15,C19)). This open a possibility to
measure the proton transversity using the {\it longitudinally} polarized
target.

The asymmetry measured by HERMES
\begin{equation}
A_{OL}=
\frac{\int{\rm d}\phi\sin\phi({\rm d}\sigma^+/{\rm d}\phi)}{P^+_H\int {\rm
d}\phi({\rm d}\sigma^+/{\rm d}\phi)}-  \frac{\int{\rm d}\phi\sin\phi({\rm
d}\sigma^-/{\rm d}\phi)}{P^-_H\int {\rm d}\phi({\rm d}\sigma^-/{\rm
d}\phi)}\ ,   \label{asym}
\end{equation}
where $P^\pm_H$ is the nucleon longitudinal polarization ($\pm$ sign means
different spin directions) averaged over transversal momenta (assuming a
Gaussian distribution) should reads as (\cite{mulddis} Eq. (115))
\begin{equation}
A_{OL}=
{2(2-y)\sqrt{1-y}({M\over Q})\sum_a e_a^2 x^2h_1^a(x)H^{\perp a/\pi^+}_1(z)/z
\over(1-y+y^2/2)\sum_a e_a^2 xf_1^a(x)D^{a/\pi^+}_1(z)}\cdot
\frac{1}{\sqrt{1+\langle p_T^2\rangle /\langle k_T^2\rangle }}
\end{equation}
where $\langle p_T^2\rangle $ and $\langle k_T^2\rangle $ are mean square of 
a transversal momenta of quark in the distribution and fragmentation 
functions.

Averaging separately numerator and denominator over $Q^2,\ y$ and $z$,
and taking into account only the u-quark distribution in the proton
($f_1^u(x)\equiv u(x)$) which gives a  dominant contribution for the $\pi^+$
production and assuming  $\langle p_T^2\rangle =\langle k_T^2\rangle $, 
one can obtain for asymmetry (\ref{asym})
\begin{equation}
\label{xdep}
A_{OL}(x)=\frac{2.1}{\langle z\rangle \sqrt2}\cdot
\left[{h_1^u(x)\over u(x)}\right]\cdot
\left[\frac{\langle H^{\perp u/\pi^+}_1(z)\rangle }{\langle D^{u/\pi^+}_1(z)\rangle }\right]
\cdot x
\end{equation}

Experimentally \cite{avakdis99}, $A_{OL}(x)$ for $\pi^+$ up to $x=0.3$
looks like a linear function
\begin{equation}
A_{OL}(x)=(0.23\pm0.06)x
\end{equation}
With $\langle z\rangle =0.41$ and  DELPHI result (\ref{apower}) this gives
an estimation for the ratio \begin{equation}
{h_1^u(x)\over u(x)}= const=0.49\pm0.18
\label{ratio}
\end{equation}

Assuming the validity of (\ref{ratio}) for valence parts in the
whole interval of $x$ one could obtain an estimation for the $u$-quark
contribution to the proton tensor charge
\begin{equation}
g^u_T\equiv \int dx\Big(h_1^u(x)-h_1^{\bar u}(x)\Big) = 0.96\pm0.36
\end{equation}
that is close to the result of the chiral quark-soliton model
\cite{pol96} $g^u_T\approx 1.12$ and to the limit followed from the density
matrix positivity constraint at $Q^2=2.5\,GeV^2$ \cite{sof98} $g^u_T\le
1.09$ .

Concerning the asymmetry observed by SMC \cite{bravardis99} on
transversely  polarized target one can state that it agrees with the 
result of HERMES.

Really, SMC has observed the azimuthal asymmetry
${\rm d}\sigma(\phi_c)\propto const\cdot(1+a\sin\phi_c)$,
where $\phi_c=\phi_h+\phi_S-\pi$ ($\phi_S$ is the azimuthal angle of the
polarization vector) is the so-called Collins angle. The raw asymmetry
$a=P_T\cdot f\cdot D_{NN}\cdot A_N$, where $P_T,\, f,$ and
$D_{NN}=2(1-y)/\left[1+(1-y)^2\right]$ are the target  polarization value,
the dilution factor and the spin transfer coefficient.

The physical asymmetry $A_N$ averaged over transverse momenta (assuming
again a Gaussian form) is given by the expression (see \cite{mulddis} Eq. 
(116))
\begin{equation}
A_N=\frac{\sum_a e_a^2 xh_1^a(x)H^{\perp a/\pi^+}_1(z)/z}
{\sum_a e_a^2 xf_1^a(x)D^{a/\pi^+}_1(z)}\cdot
\frac{1}{\sqrt{1+\langle p_T^2\rangle /\langle k_T^2\rangle }}
\end{equation}

Integration over $x$ and $z$, and using again the approximation of the 
$u$-quark dominance and $\langle p_T^2\rangle =\langle k_T^2\rangle $ gives
from the experimental value $A_N=0.11\pm0.06$, $\langle z\rangle =0.45$ and 
DELPHI result
(\ref{apower})
\begin{equation}
{\langle xh_1^u(x)\rangle \over\langle xu(x)\rangle }=0.54\pm0.35
\label{ratio2}
\end{equation}
what in agreement with the HERMES ratio (\ref{ratio}).

\section{Conclusions}

In conclusion, I would like to stress that there are several ways allowing
one to measure the transverse quark polarization among which the use of the
T-odd PFF looks like the most perspective for future experiments in
measuring of transversity, like COMPASS at CERN. I present the first
experimental estimation for the absolute value of analyzing power of the
method. Of course, a more accurate measurements of it is necessary.
However, even now it allows the first crude estimation of the proton
transversity from observed azimuthal asymmetry in  semi-inclusive DIS. The
most interesting discovery here is a good agreement of transversities
obtained from transversally and longitudinally polarized targets due to
small contribution of $h_{1L}^\perp(x)$. This allows measuring
the transversity in the same experiments as for $\Delta g$.

\medskip
{\footnotesize I would like to thank K.Goeke, M.Polyakov and the Institute
for Theoretical Physics II of Ruhr University Bochum, where part of this
work was done, for discussions and warm hospitality.}


\end{document}